\documentclass[journal]{IEEEtran}
\ifCLASSINFOpdf
\else
\fi
\usepackage{graphicx}
\usepackage{mathtools}
\usepackage{booktabs}
\usepackage{multirow}
\begin{document}
%
\title{Cappuccino: Efficient Inference Software Synthesis for Mobile System-on-Chips}
%
%
%

\author{Mohammad~Motamedi, Daniel~Fong, and Soheil Ghiasi\\ Electrical and Computer Engineering Department, University of California, Davis, USA.\vspace{-1	cm}}
\markboth{}%
{Shell \MakeLowercase{\textit{et al.}}: Bare Demo of IEEEtran.cls for IEEE Journals}
%


\maketitle
\begin{abstract}
Convolutional Neural Networks (CNNs) exhibit remarkable performance in various machine learning tasks. As sensor-equipped Internet of Things (IoT) devices permeate into every aspect of modern life, the ability to execute CNN inference, a computationally intensive application, on resource constrained devices has become increasingly important. In this context, we present Cappuccino, a framework for synthesis of efficient inference software targeting mobile System-on-Chips (SoCs). We propose techniques for efficient parallelization of CNN inference targeting mobile SoCs, and explore the underlying tradeoffs. Experiments with different CNNs on three mobile devices demonstrate the effectiveness of our approach.
\vspace{-0.5cm}
\end{abstract}

\IEEEpeerreviewmaketitle

\section{Introduction}

%
\IEEEPARstart{C}{onvolutional} Neural Networks (CNNs) have proven to be one of the most effective approaches to feature extraction \cite{krizhevsky2012imagenet,szegedy2015going}. While frameworks such as Caffe \cite{jia2014caffe} or Torch \cite{collobert2011torch7} are commonly used for training CNN models, inference using trained CNNs on resource-constrained platforms remains a challenge. We hypothesize that platforms based on mobile system-on-chips (SoCs) will be a major player in the emerging IoT landscape due to their rich feature set and market forces, and thus, we contend that efficient CNN inference on such platforms is increasingly essential. 

Forward evaluation of a trained CNN, also known as inference, is computationally intensive. The research community has put forth a number of solutions for accelerating CNN inference on different platforms, including the design of a customized ASIC chip (\cite{chen2014dadiannao,jouppi2016google}), FPGA-based accelerator design (\cite{zhang2015optimizing,motamedi2016design,qiu2016going}), and parallelization on server-grade graphics processing units (GPUs). Latifi \emph{et al.} offered a library for parallel execution of CNNs on mobile devices~\cite{latifi2016cnndroid}. 

We present Cappuccino, a tool for automatic synthesis of efficient CNN inference software targeting mobile SoCs. In addition to the software synthesis capability, Cappuccino features a novel approach to zero-overhead utilization of vector instructions. Furthermore, it considers the effect of inexact computing on classification accuracy, and leverages imprecise arithmetic to further optimize the computation.

\section{Convolutional Neural Networks}
\label{SEC:CNN}
\begin{figure}
	\includegraphics[width=\columnwidth]{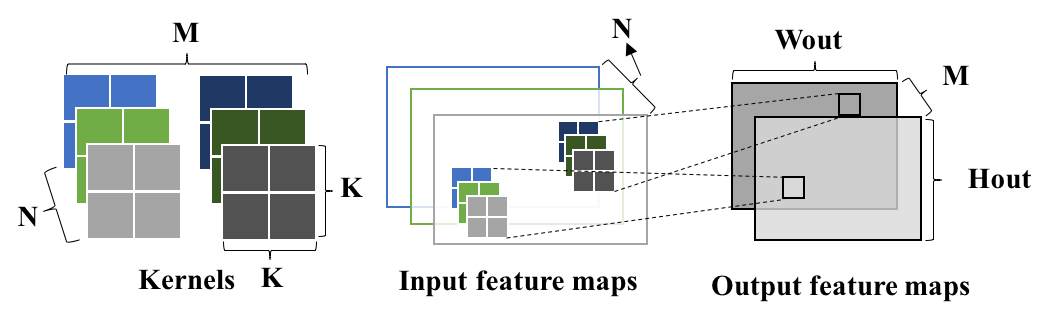}
	\centering
	\caption{A convolutional layer. $M$ filter banks are convolved with $N$ IFMs to generate $M$ OFMs.}
	\label{fig:conv}
	\vspace{-0.2cm}
\end{figure}

\begin{figure}
	\includegraphics[width=7cm]{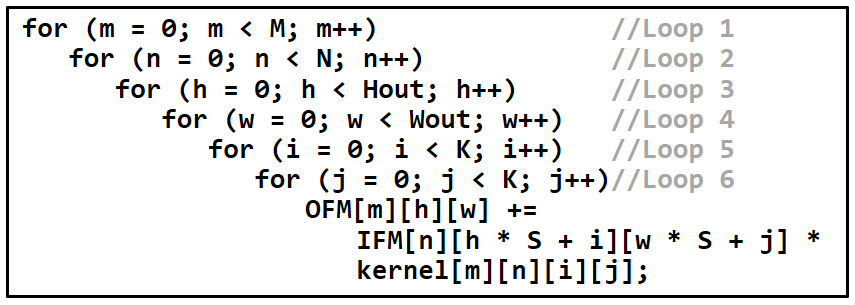}
	\centering
	\caption{A simplified pseudo-code for a convolutional layer. The computation involves six nested loops, which results in a high computational complexity.}
	\label{fig:conv_code}
	\vspace{-0.4cm}
\end{figure}

Modern Convolutional Neural Networks (CNNs) have millions of parameters, whose values are obtained during training.
Each CNN has multiple convolutional layers, which use 3D filter banks for feature extraction. The convolution result of kernels of a filter bank with Input Feature Maps (IFMs) is accumulated to create Output Feature Maps (OFMs). The number of IFMs, the number of OFMs, and the output size are $N$, $M$, and $Wout \times Hout$, respectively. The convolution operation is visualized in Figure \ref{fig:conv}.
Thus, a CNN layer has $M \times N$ kernels ($M$ filter banks with $N$ kernels each). Kernels have dimension of $K \times K$.
Each pixel in an OFM is the sum of convolutions between kernels and the corresponding pixels in IFMs. To generate adjacent pixels in an OFM, the kernel bank is slid across IFMs by a stride of $S$. A simplified pseudo-code for a convolution operation is shown in Figure \ref{fig:conv_code}. The vast majority of CNN inference execution time is spent in convolutional layers \cite{cong2014minimizing}, and thus, we restrict our discussion to them.
\vspace{-0.4cm}
\section{Cappuccino}
\label{SEC:CAP}
\begin{figure*}
	\includegraphics[width=\textwidth]{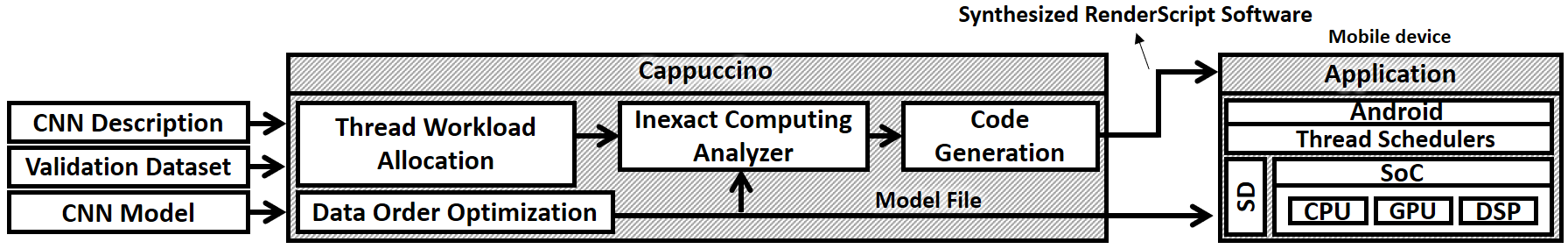}
	\centering
	\caption{Primary Program Synthesizer creates a primary parallel program. Cappuccino uses this program to determine in what layers which inexact computing mode can be used. Finally, the software synthesizer uses the results of this analysis to generate RenderScript-based inference software.	
	}
	\vspace{-0.5cm}
	\label{fig:Cappuccino}
\end{figure*}
In order to use a CNN for inference on mobile devices, one has to evaluate its forward path with known parameter values. Cappuccino serves this very purpose in that, it synthesizes an optimized SoC-based inference software for a given CNN description. Our current embodiment of Cappuccino synthesizes the CNN in form of an optimized RenderScript program, which exploits the available processing resources on a mobile SoC to execute the computation. Depending on the target SoC, the generated program will be typically launched on multiple CPU cores, the mobile GPU, and the mobile DSP. 
 
As Figure \ref{fig:Cappuccino} illustrates, Cappuccino requires three inputs. The first is a network description file that contains the CNN architectural information such as number, size, and type of its layers. The second input is a model file, which contains the weight and bias parameter values. Cappuccino reorders CNN parameters to improve the performance of vectorized operations. Parameter reordering does not change the model size, and occurs during compile-time. The third input is the validation dataset that was originally used during training of the CNN. Using this dataset, Cappuccino analyzes the impact of optimizations, such as inexact computing on the given CNN, to determine the suitability of utilizing imprecise arithmetic for the given CNN. The result of this analysis guides the corresponding decision during software generation. 


\section{Inference Optimization Strategies}
\label{SEC:ACC}
\subsection{Thread Workload Allocation}
Convolutional layers contain three main sources of parallelism: Kernel-Level Parallelism (KLP), Filter bank-Level Parallelism (FLP), and Output-Level Parallelism (OLP). In accelerating a CNN, one or more of these types of parallelism should be used for workload allocation to threads. 
\subsubsection{Kernel-Level Parallelism (KLP)}
In KLP, parallelization is obtained by executing the computations for convolving a kernel with corresponding IFM pixels in parallel. Hence, each thread computes one multiplication, and the final result is generated via accumulation by an eventual reduction operation.
\subsubsection{Filter bank-Level Parallelism (FLP)}
In FLP, filter banks exploit parallelism by allocating kernel computation to separate threads. In this case, each thread computes the convolution of an entire kernel. Subsequently, a reduction addition yields the final result.
\subsubsection{Output-Level Parallelism (OLP)}
In OLP, computation of output pixels are carried out in parallel. That is, each software thread computes the 3D convolution of an entire filter bank of $N$ kernels and corresponding pixels in IFMs.

An advantage of OLP is that a kernel loaded by one thread can be reused by other threads that are responsible for generating another pixel in the same OFM. In SoCs with efficient cache systems, it is possible to load each kernel once and use it $Wout \times Hout$ times. In contrast, data loaded in KLP and FLP cannot be reused as efficiently by the same or other threads. Moreover, in KLP and FLP the required reduction incurs additional overhead for thread synchronization and inter-thread data transfer. As such, Cappuccino uses OLP as its primary workload allocation policy at the thread level. Furthermore, it utilizes vector processing to exploit KLP and FLP within each thread.

\begin{figure}
	\includegraphics[width=7cm]{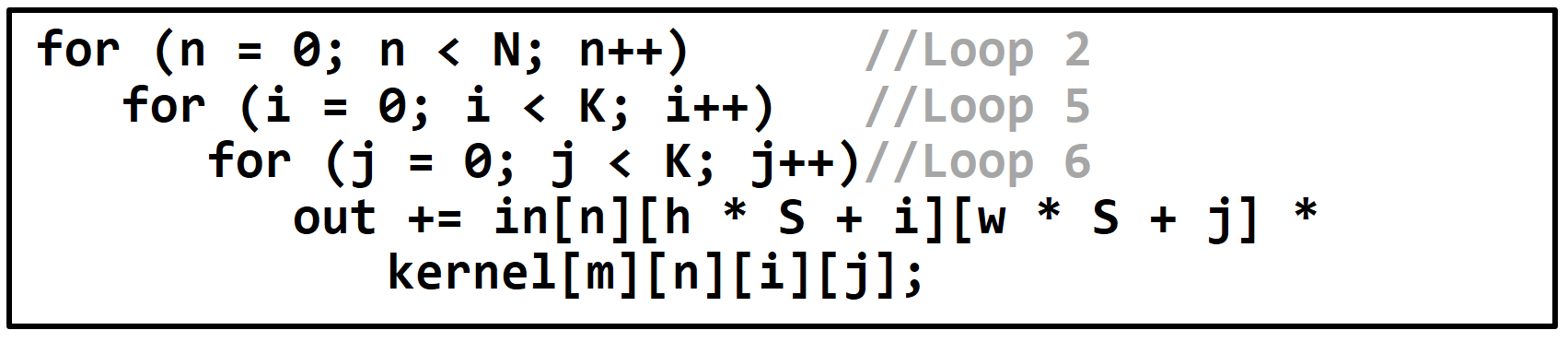}
	\centering
	\caption{Workload allocated to a thread. Values of $w, h$, and $m$ can be computed using the thread Id.}
	\label{fig:code1}
	\vspace{-0.6cm}
\end{figure}

The output of each convolutional layer is a 3D data structure which includes $\alpha = M \times Wout \times Hout$ elements (also referred to as pixels). These elements can be uniquely identified using three variables: the OFM ($m$), the column ($w$), and the row ($h$) number. Each element is the result of a convolution between the corresponding window of an IFM and a specific filter bank (Figure \ref{fig:code1}). Each thread is identified using a unique index $x$, where  $x \in [0,~\alpha)$. The identifier index is used to compute values of $w$, $h$, and $m$.  
\subsection{Data Reordering for Vector Processing}
\begin{figure}
	\includegraphics[width=7cm]{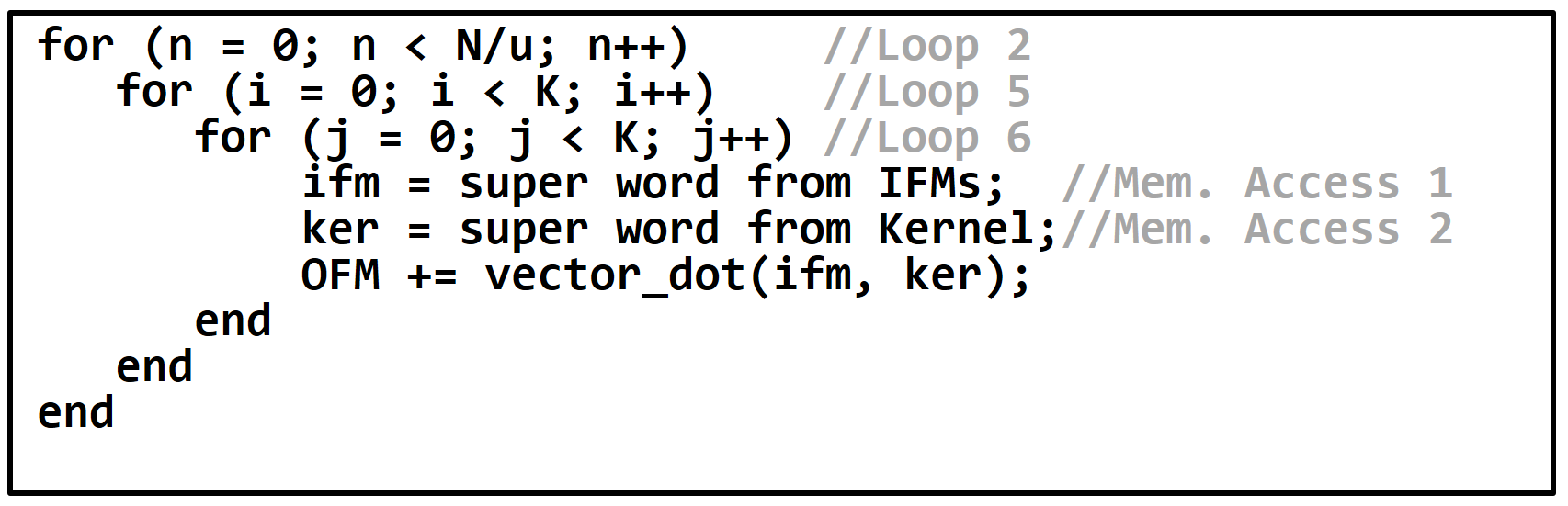}
	\centering
	\caption{Cappuccino optimizes data ordering to enable efficient vector processing. $u$ words are loaded per access. Subsequently, a vectorized MAC operation performs partial convolution on $2u$ operand elements in parallel.}
	\label{fig:code2}
	\vspace{-0.5cm}
\end{figure}

Cappuccino uses vector processing to further optimize intra-thread workload execution. Before executing a vector instruction, it is necessary to load all of the operands. In most SoCs with vector processing support, the memory bus is wide enough to load multiple words of a contiguous block in one memory access. 
To utilize this feature, known as memory access locality, model parameter values have to be shuffled around.
Conventionally, IFMs and kernel parameters are stored in either row- or column- major order. Therefore, data elements stored in the adjacent memory addresses are either the next element from the same row/column or the first element of the next row/column. If we represent an element's address using (Layer, Row, Column) format, the data stored in a row major format reads:
\begin{equation}
\footnotesize (0, 0, 0), (0, 0, 1), \cdots, (0, 1, 0), (0, 1, 1), \cdots, (0, 2, 0), (0, 2, 1), \cdots
\label{EQ: row}
\end{equation}
In a $u$-way vector processor, one wants to load at least $u$ operands with a single memory access. Cappucino reorders the model data to achieve this goal. In particular, we propose to store the model data in a map major order, as opposed to row or column major, so that a thread can apply vector instructions to corresponding elements of different maps. Absent of this optimization, vector processing would incur significant overhead at the boundaries of a kernel. For example, assuming $u = 4$, we reorder the model data in the following order (\ref{EQ: vectorized}): 
\begin{equation}
\label{EQ: vectorized}
\footnotesize
\begin{split}
(0, 0, 0), (1, 0, 0), (2, 0, 0), (3, 0, 0), (0, 0, 1), (1, 0, 1), (2, 0, 1), \\ (3, 0, 1), \cdots, (4, 0, 0), (5, 0, 0), (6, 0, 0), (7, 0, 0), \cdots
\end{split}
\end{equation}
\begin{figure}
	\includegraphics[width=7cm]{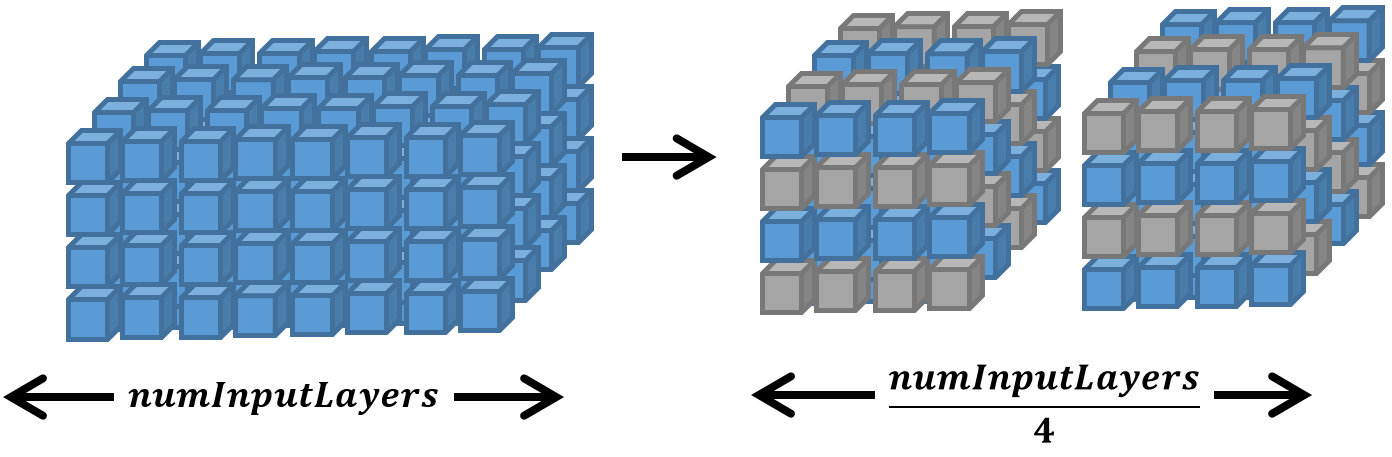}
	\centering
	\caption{Efficient vector processing enabled by data reordering from either row or column major to map major. Groups of $u$ Elements, in either gray or blue, in the same spatial location from consecutive feature maps form a vector (assuming $u = 4$).}
	\label{fig:reshape}
	\vspace{-0.5cm} 
\end{figure}
A 3D representation of this transform is shown in Figure \ref{fig:reshape}. When model data is reordered, Cappuccino reads IFMs as super-words (vectors), performs vectorized convolution, and accumulates the result. The optimized computation is shown in the algorithm of Figure \ref{fig:code2}.
\subsubsection{Zero-Overhead Dynamic Reordering of OFMs}
\begin{figure}
	\includegraphics[width=5cm]{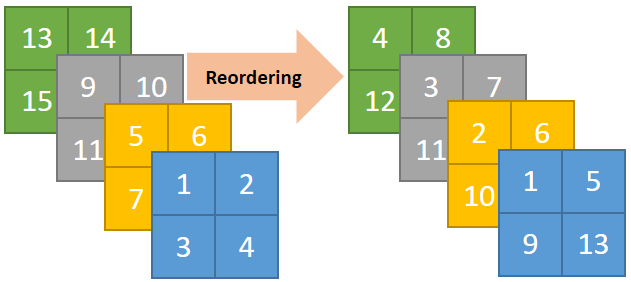}
	\centering
	\caption{Zero overhead reordering of OFMs: Cappuccino directly stores data in the map major, in lieu of row major format. The case of $u = 4$ is illustrated.}
	\label{fig:data_reshape2}
	\vspace{-0.5cm} 
\end{figure}

Note that model data can be reordered and written to a new model file without any overhead as it happens statically at compile-time. However, reordering the input to an intermediate CNN layer is not as straight forward. In CNNs, the output of a layer becomes the input to the next layer. It follows that the output of a layer has to be reordered to allow the use of vectorized operations in  computing the next CNN layer. This process has to happen dynamically, and thus, is expected to incur time and energy overhead. 

Cappuccino avoids the dynamic data reordering overhead by directly storing elements of the OFMs in map major order as they are computed. Parameters $w$, $h$, and $m$ are used to determine the location of the output element that thread $x$ generates. To store OFMs in map major format, one has to swap the priorities associated with these parameters. For example, the result of computations by the second thread ($x = 1$) is by default stored in the second location of the output memory. After reordering, however, the second element of the output memory must contain $(m = 1, h = 0, w = 0)$. Such an output can be directly used as the input to the next layer without any overhead. Figure \ref{fig:data_reshape2} illustrates the idea.

To create the output in the reordered map major format, we generate indexes for stacks of $u$ layers, instead of a single layer (Figure \ref{fig:data_reshape2}). That is, we start indexing the second row only after all first rows of all $u$ layers are indexed. Equations (\ref{EQ: w_new}) and (\ref{EQ: h_new}) map a thread id to $w$ and $h$, respectively. For computing the value of $m$ (map index), it is required to see which stack and layer a particular output belongs to (Figures \ref{fig:reshape} and Figure \ref{fig:data_reshape2}). Equation (\ref{EQ: m_new}) computes the value of $m$.  
\begin{equation}
\footnotesize w = \lfloor x / u\rfloor~\%~Wout
\label{EQ: w_new}
\end{equation}
\begin{equation}
\footnotesize  h = \lfloor\frac{x}{u \times Wout}\rfloor~\%~Hout
\label{EQ: h_new}
\end{equation}
\begin{equation}
\footnotesize m = (x~\%~u) + \lfloor\frac{x}{u \times Wout \times Hout}\rfloor \times u
\label{EQ: m_new}
\vspace{-0.4cm} 
\end{equation}
\subsection{Inexact Computing}
Modern mobile SoCs tend to support a number of predefined imprecise computing modes that are likely to results in faster or more energy efficient execution~\cite{mitra2013use}. On such platforms, the target processing mode has to strike a balance between the implementation metrics, e.g., runtime or energy dissipation, and the inference classification accuracy. 

For example, RenderScript offers two imprecise computing modes for applications that do not need a strict implementation of the IEEE 754 standard, called relaxed and imprecise computing modes. In both modes, the implementation of floating point arithmetic is not fully compliant with the IEEE standard 754 for handling denormalized numbers.  The imprecise computing mode is more efficient, but has a lower arithmetic accuracy. In this mode, operations resulting in -0.0 can return +0.0, and operations on INF and NAN are unsupported. Perhaps more importantly, vector processing is only available under imprecise computing modes, in current version of RenderScript. Vector processing under the RenderScript precise computing mode would result in sequential processing of vector elements. 

Cappuccino analyzes the given CNN layer by layer to determine the best matching computing mode for every layer of the CNN. In every layer, it utilizes the validation dataset to measure the classification accuracy under different processing modes. Subsequently, Cappuccino determines which layers of a CNN can be processed using inexact arithmetic and which ones demand a precise implementation. The goal is to execute as many CNN layers as possible in inexact modes, under user specified constraints in terms of acceptable degradation in classification accuracy.

In our discussions, the term accuracy arises in reference to either arithmetic accuracy or classification accuracy. The former measures the numerical difference between values computed in exact vs. inexact arithmetic. The latter indicates the inference classification performance of CNNs, for example by measuring the percentage of true positive predictions. 
\vspace{-0.4cm} 
\section{Experimental Results}
\label{SEC:EXP}
\subsection{Setup}
\begin{table*}
	\centering
	\caption{Experimental results with three modern CNNs on three different platforms. Execution times are reported in milliseconds. Baseline is a single-threaded implementation.}
	\scalebox{0.9} {
		\begin{tabular}{ccccccccccccc}
			\toprule[.07cm]
			\multirow{2}{*}{CNN Name}                           &        \multicolumn{4}{c}{Execution Time on Nexus 5 (ms)}        &       \multicolumn{4}{c}{Execution Time on Nexus 6P (ms)}        &       \multicolumn{4}{c}{Execution Time on Galaxy S7 (ms)}       \\
			\cmidrule[.05cm](l){2-5} \cmidrule[.05cm](l){6-9} \cmidrule[.05cm](l){10-13} & Baseline & Parallel & Imprecise & Speedup & Baseline & Parallel & Imprecise & Speedup & Baseline & Parallel & Imprecise & Speedup \\ \toprule[.07cm]
			AlexNet                                    & 33848.40 &  947.15  &  836.32   & 40.47X  &   8626   &  512.72  &   61.80   & 139.58X & 8698.43  &  442.97  &  127.78   & 68.07X  \\ \midrule
			SqueezeNet                                  & 43932.73 & 1302.10  &  161.50   & 272.03X & 17299.55 &  671.46  &  141.30   & 122.43X & 12331.82 &  888.91  &  150.24   & 82.08X  \\ \midrule
			GoogLeNet                                   & 84404.40 & 2651.12  &  2478.09  & 34.06X  & 25570.48 & 1575.45  &  602.28   & 42.46X  & 21917.67 & 1699.42  &  686.08   & 31.95X  \\ \bottomrule[.07cm]
			\vspace{-0.8cm}
		\end{tabular}
	}
	\label{TBL:EXE}	
\end{table*}
We used Cappuccino to implement three modern CNNs: AlexNet \cite{krizhevsky2012imagenet}, GoogLeNet \cite{szegedy2015going}, and SqueezeNet \cite{iandola2016squeezenet}. Subsequently, the parallelized implementations are evaluated on three different smartphones with different generations of Qualcomm Snapdragon SoCs. 
In order to increase the precision in measurements, all experiments have been repeated 100 times, the minimum and maximum observations are omitted, and the average of the remaining 98 observations are reported. In all of the experiments, the smartphones were put in airplane mode, their screen brightness were fully dimmed, and their background processes were stopped to the extent possible.
\vspace{-0.5cm}
\subsection{Runtime and Energy Efficiency}
\begin{table}
	\centering
	\caption{Energy Consumption in joules. Baseline is a single-threaded program.}
	\scalebox{0.7} {
		\begin{tabular}{cccccccc}
			\toprule[.07cm]
			            \multirow{2}{*}{CNN Name}             &  \multicolumn{3}{c}{Baseline (J)}  &  \multicolumn{3}{c}{Proposed (J)}  & \multirow{2}{*}{Ratio} \\
			\cmidrule[.05cm](l){2-4} \cmidrule[.05cm](l){5-7} & First 1000 & Second 1000 & Average & First 1000 & Second 1000 & Average &  \\ \toprule[.07cm]
			                   SqueezeNet                     &   26.37    &    26.40    &  26.39  &   3.39    &    3.36    &  3.38  &         7.81X         \\ \bottomrule[.07cm]
		\vspace{-1cm}
		\end{tabular}
	}
	\label{TBL:ENG}	
\end{table}	

\subsubsection{Speedup}
We executed the synthesized programs on the platforms and measured the execution time. Table \ref{TBL:EXE} summarizes the results. Programs synthesized by Cappuccino offer a speedup of at least 31.95X (GoogLeNet on Galaxy S7) and at most 272.03X (SqueezeNet on Nexus 5) compared to the baseline implementation of single-threaded Java. Moreover, the execution time in all but one case is below a second. 
\subsubsection{Effect of Inexact Computing}
To determine the best inexact computing mode, we use Cappuccino to measure the classification accuracy of the aforementioned CNNs in computing modes supported by target platforms. This analysis is performed on 5000 random images of ILSVRC 2012 validation dataset \cite{ILSVRC15}. The classification accuracy in imprecise mode turns out to be identical to the exact mode. Hence, Cappuccino recommends utilization of imprecise computing in all layers.

Table \ref{TBL:EXE} demonstrates the effect of imprecise computing on execution time. In our experiments, use of imprecise computing mode offers up to 8X speedup compared to the same implementation under exact arithmetic. Note that RenderScript incarnation of the imprecise computing mode enables vector processing in addition to other optimizations, such as using a rapid exception handling for denormalized numbers.
\subsubsection{Comparison with Related Work}
Table \ref{TBL:Comparison} compares the performance of software synthesized by Cappuccino with the state-of-the-art work \cite{latifi2016cnndroid}. The proposed solution under exact arithmetic improves the execution time by 1.38X. In addition, when the synthesized software is both parallel and imprecise, it shows up to 11.47X speedup compared to CNNDroid \cite{latifi2016cnndroid}.
\subsubsection{Energy Consumption} Cappuccino invokes many threads, which increases the instantaneous power consumption compared to a sequential program. However, software synthesized by Cappuccino runs drastically faster than a sequential equivalent. This results in reduction of energy consumption. Table \ref{TBL:ENG} compares the energy consumption for running SqueezeNet on Nexus 5. Reported numbers are computed by running each program 1000 times, and calculating the average. Measurements are performed twice to showcase repeatability (2000 runs total).
\vspace{-0.3cm}
\section{Conclusion}
\label{SEC:CON}
\begin{table}
	\centering
	\caption{Performance of Cappuccino versus prior art (CNNDroid)~\cite{latifi2016cnndroid} for running AlexNet on Qualcomm Snapdragon 810.}
	\scalebox{0.65} {
		\begin{tabular}{cccccc}
			\toprule[.06cm]
			& CNNDroid \cite{latifi2016cnndroid} & Cappuccino: Parallel &  Speedup &      Cappuccino: Imprecise   & Speedup      \\ \toprule[.07cm]
			Execution Time (ms) &                709                 &            512.72 &      1.38X       &  61.80&      11.47X    \\ \bottomrule[.07cm]
		\vspace{-0.8cm}
		\end{tabular}
	}
	\label{TBL:Comparison}	
\end{table}
In this paper we presented Cappuccino, a platform for efficient synthesis of CNN inference software targeting mobile SoCs. Cappuccino leverages RenderScripts via which, it utilizes CPUs, the GPU and the DSP that commonly exist on a mobile SoC to execute a CNN efficiently. Cappuccino performs an assessment on the impact of inexact computing on execution time and classification accuracy. Subsequently, it selects an inexact computing mode that best fits a layer of a CNN. Compared to sequential implementations, programs synthesized by Cappuccino achieve a speedup of at least 31.95X and at most 272.03X, and improve the energy consumption by 7.81X.
\ifCLASSOPTIONcaptionsoff
\newpage
\fi


%

\bibliographystyle{IEEEtran}
\bibliography{IEEEabrv,ESL}

%





\end{document}